\documentclass[aps,prl,reprint,superscriptaddress,showpacs,floatfix]{revtex4-1}

\usepackage[draft]{changes}
\usepackage[normalem]{ulem}
\usepackage{graphicx}
\usepackage{amssymb} 
\usepackage{dcolumn}
\usepackage{bm}
\usepackage[mathlines]{lineno}
\usepackage[colorlinks,linkcolor=blue,anchorcolor=blue,citecolor=blue,urlcolor=blue]{hyperref}
\usepackage{multirow}
\usepackage{color}
\usepackage{amsmath}
\usepackage[normalem]{ulem}

\makeatletter
\renewcommand*{\@fnsymbol}[1]{\ensuremath{\ifcase#1\or \dagger\or *\or \ddagger\or
\mathsection\or \mathparagraph\or \|\or **\or \dagger\dagger \or
\ddagger\ddagger \else\@ctrerr\fi}} \makeatother

\usepackage{color}

\usepackage[draft]{changes}

\usepackage{color}

\begin{document}

\title{Anomalous thermal conductivity in 2D silica nanocages of immobilizing noble gas atom} 

\author{Yang Wang}
\affiliation{State Key Laboratory for Mechanical Behavior of Materials, School of Materials Science and Engineering,
Xi'an Jiaotong University, Xi'an 710049, China}

\author{Zhibin Gao}
\thanks{Authors to whom correspondence should be addressed: \url{zhibin.gao@xjtu.edu.cn}, and \url{y.zhao@ytu.edu.cn}}
\affiliation{State Key Laboratory for Mechanical Behavior of Materials, School of Materials Science and Engineering,
Xi'an Jiaotong University, Xi'an 710049, China}

\author{Xiaoying Wang}
\affiliation{State Key Laboratory for Mechanical Behavior of Materials, School of Materials Science and Engineering,
Xi'an Jiaotong University, Xi'an 710049, China}

\author{Jinping Sun}
\affiliation{School of Materials Science and Engineering, Harbin Institute of Technology at Weihai, 2 West Wenhua Road, Weihai 264209, China}

\author{Minxuan Feng}
\affiliation{State Key Laboratory for Mechanical Behavior of Materials, School of Materials Science and Engineering,
Xi'an Jiaotong University, Xi'an 710049, China}

\author{Yuzhou Hao}
\affiliation{State Key Laboratory for Mechanical Behavior of Materials, School of Materials Science and Engineering,
Xi'an Jiaotong University, Xi'an 710049, China}

\author{Xuejie Li}
\affiliation{State Key Laboratory for Mechanical Behavior of Materials, School of Materials Science and Engineering,
Xi'an Jiaotong University, Xi'an 710049, China}

\author{Yinchang Zhao}
\thanks{Authors to whom correspondence should be addressed: \url{zhibin.gao@xjtu.edu.cn}, and \url{y.zhao@ytu.edu.cn}}
\affiliation{Department of Physics, Yantai University, Yantai 264005, China}

\author{Xiangdong Ding}
\affiliation{State Key Laboratory for Mechanical Behavior of Materials, School of Materials Science and Engineering,
Xi'an Jiaotong University, Xi'an 710049, China}



\begin{abstract}
Noble gas atoms, such as Kr and Xe are byproducts of nuclear fission in nuclear plants. How to trap and confine these volatile even radioactive gases is particularly challenging. Recent studies have shown that they can be trapped in nanocages of ultrathin silica. Here we exhibit with self-consistent phonon (SCP) theory and four-phonon (4ph) scattering where the adsorption of noble gases results in an anomalous increase in lattice thermal conductivity ($\kappa_L$), while the presence of Cu atoms doping leads to a reduction in $\kappa_L$. We trace this behavior in host-guest 2D silica to an interplay of tensile strain, rattling phonon modes, and redistribution of electrons. We also find that 4ph scatterings play indispensable roles in $\kappa_L$ of 2D silica. Our work illustrates the microscopic heat transfer mechanism in 2D silica nanocages with the immobilization of noble gas atoms and inspires further exploring materials with the kagome and glasslike $\kappa_L$.
\end{abstract}

\maketitle
Silicon dioxide (silica), one of the most abundant materials in the 
earth's cruest, constitutes a fundamental component of glass, sand, 
and the majority of minerals. With the advancement of nanomaterial 
fabrication technology, single-crystal two-dimensional (2D) silica 
films, using chemical vapor deposition (CVD), can be successfully 
grown on metal Mo(112)~\cite{PhysRevLett.95.076103}, 
Ru(0001)~\cite{PhysRevLett.105.146104}, and even graphene 
substrate~\cite{doi:10.1021/nl204423x,doi:10.1126/science.1242248}. 
This thinnest insulating material exhibits distinctive features~\cite{doi:10.1021/acs.nanolett.6b03921}, 
and are also useful
in confined chemical reaction~\cite{https://doi.org/10.1002/anie.201802000},
subatomic species transport~\cite{doi:10.1126/science.abd7687}, and 
radioactive gas separation~\cite{https://doi.org/10.1002/smll.202103661}.
Moreover, atomically smooth 2D silica can be mechanically exfoliated and transferred to the support at millimeter scale~\cite{doi:10.1021/acsnano.6b03929}.
This property renders it suitable for catalysis and the isolation of 
graphene from metal substrates, facilitating the creation of vertically 
transferable heterostructures~\cite{doi:10.1021/acs.nanolett.7b02776,doi:10.1021/nl902558x,LIU2016203} 
and ultrathin gate oxides in field effect transistors~\cite{muller1999electronic}.

Radioactive isotopes of noble gases Krypton (Kr) and Xenon (Xe) are a byproduct of nuclear fission in nuclear plants~\cite{kawai2016van,ncomms16118,https://doi.org/10.1002/adfm.201806583}.
Compared to electrostatic trapping, the only approach to trap noble gas at room temperature is the ion implantation~\cite{dil2008surface,cun2013immobilizing}, forming a 2D silicate-noble gas clathrate compounds~\cite{ncomms16118,https://doi.org/10.1002/adfm.201806583,https://doi.org/10.1002/smll.202103661}.

Introducing guest atoms in host-guest systems like clathrate can effectively reduce their $\kappa_L$. 
Concurrently, the rattling modes derived from guest atoms significantly enhance the scattering phase space and reduce phonon relaxation time ($\tau$), resulting in a decrease in $\kappa_L$~\cite{ZHANG2018289,li2015ultralow}. Therefore, the application of 2D silica nanocage for capturing fission gases may potentially further 
amplify temperature gradients within nuclear fuels, compromising safety. 

In the recently scrutinized lead-phosphate crystal Pb$_{10}$(PO$_4$)$_6$O (LK-99), which underscores the pivotal role played by Cu atoms in the modulation of electronic properties~\cite{lai2024first}. Noble gas atoms possess saturated electron structures, and doping noble gas atoms can yield more idealized models. Hence, contrasting host-guest systems doped with Cu atoms and noble gas atoms allow for separate consideration of the mechanism by which electron density distribution and phonon-phonon scatterings~\cite{Tang_2021}.

In this study, we investigated the microscopic heat transfer mechanism in the 2D silica nanocages. 
The results indicate contrasting effects of adsorbed noble gas and Cu atoms on the $\kappa_L$ of 2D silica. On one hand, the adsorption of the Kr atom induces tensile 
strain in the host system, decreasing phonon scattering probabilities and resulting in an increase in $\kappa_L$. On the other hand, Cu atom doping 
confines the electronic distribution of the 
nanocage and suppresses strain in the host system. Moreover, substantial flat bands emerge in the low-frequency acoustic phonon branch, enhancing phonon-phonon scattering through rattling modes, leading to a reduction in $\kappa_L$. 

DFT calculations were performed using the Vienna ab initio simulation package (VASP)~\cite{PhysRevB.54.11169}. The interaction between valence electrons and ions is realized by the functional of Perdew-Burke-Ernzerhof (PBE)~\cite{PhysRevLett.77.3865} and the generalized gradient approximation (GGA) with projector augmented wave (PAW)~\cite{PhysRevB.59.1758}. The format of the second-order interatomic force constants (2nd-IFCs) file from the Alamode~\cite{PhysRevB.92.054301} was transformed into the ShengBTE interface~\cite{PhysRevB.104.224304}.
For three-phonon (3ph) and 4ph scatterings, we used a uniform 40 $\times$ 40 $\times$ 1 and 7 $\times$ 7 $\times$ 1 q-mesh grids~\cite{HAN2022108179}. More details are in the Supplementary Material. 

\begin{figure*}
\includegraphics[width=0.8\textwidth]{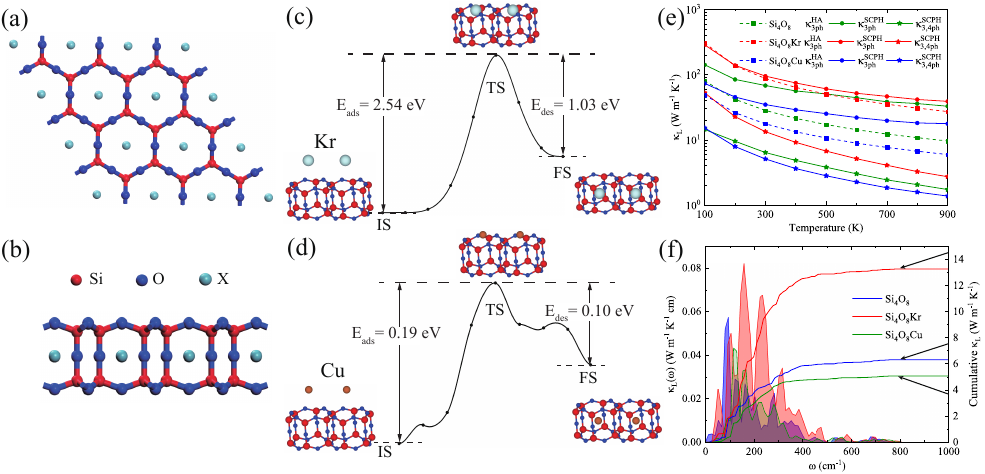}
\caption{%
(a)-(b) The top and side views of the 2D-Si$_4$O$_8$ crystal structure with doped X atoms (X=Ar, Kr, Xe, Cu). The red, blue, and cyan colors represent silicon (Si), oxygen (O), and doped atoms X. (c)-(d) The minimum energy path for doping Kr and Cu atoms.
(e) $\kappa_{3ph}^{HA}$, $\kappa_{3ph}^{SCPH}$ and $\kappa_{3,4ph}^{SCPH}$ as a function of temperature. Si$_4$O$_8$, Si$_4$O$_8$Kr, and Si$_4$O$_8$Cu represent undoped, Kr-doped, and Cu-doped silica.
(f) The $\kappa_L$ spectrum of the frequency distribution in the shaded areas and cumulative $\kappa_L$ as a function of frequency in solid lines.
\label{fig1}}
\end{figure*}

Top and side views of 2D silica with P6/\emph{mmm} symmetry are depicted in Fig.~\ref{fig1} (a-b).
The primitive cell consists of 4 silicon atoms and 8 oxygen atoms with an optimized lattice constant
of a = b = 5.312 \AA. 

Due to the minimization of structure energy, the doping atom precisely stays at the center of the nanocage to keep stability. 
We computed the phonon spectra for a series of atom dopings, as shown in Fig. S1-S2.
We find that the large electronegativity difference between the guest atoms and the host lattice could lead to structural instability since excess dopant electrons will cause a collapse of the nanocage.
Furthermore, noble gas atoms possess a saturated electron orbit and can also be stable on the framework electron structure of 2D-Si$_4$O$_8$~\cite{C6EE00322B}.

The dynamic stability of structures with phonon spectrum is further assessed. 
It is worth noting that the adsorbed noble
gas atoms results are stable isotopes in the calculation.
The minimum energy path for doping atoms are calculated using climbing image nudged elastic band (CI-NEB) calculations~\cite{10.1063/1.1329672}.
The activation energies for the adsorption and desorption of Ar, Kr, Xe, and Cu doping atoms within the nanocage are shown in Fig.~\ref{fig1} (c)-(d) and Fig. S3. 
This indicates that adsorbed atoms can be captured and released into the nanocage with proper activation energies of $E_{ads}$ and $E_{des}$.
Due to the similar properties exhibited by noble gases, in the following, we examine variations in $\kappa_{L}$ using Kr and Cu atoms as examples.

\begin{figure*}
\includegraphics[width=0.75\textwidth]{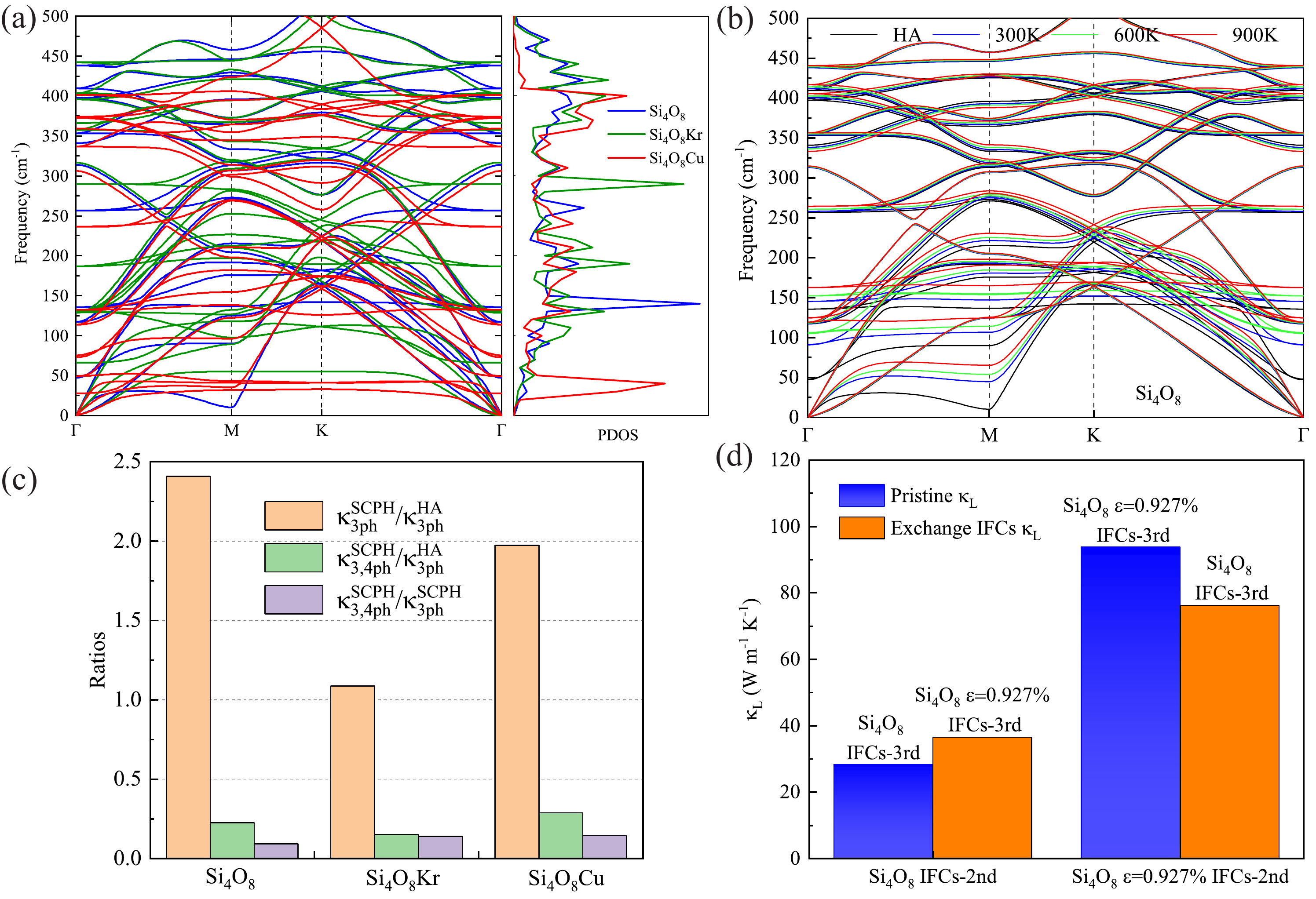}
\vspace{-2mm}
\caption{%
(a) Phonon dispersions and phonon density of states (PDOS). (b) The renormalized phonon dispersion of Si$_4$O$_8$ at different temperatures. 
(c) The ratio of $\kappa_{3ph}^{SCPH}/\kappa_{3ph}^{HA}$, $\kappa_{3,4ph}^{SCPH}/\kappa_{3ph}^{HA}$, $\kappa_{3,4ph}^{SCPH}/\kappa_{3ph}^{SCPH}$ at 300 K. (d) $\kappa_L$ of 2D silica without strain and $\varepsilon = 0.927\%$ strain calculated by the exchange IFCs at 300 K. The abscissa is the 2nd-IFCs and the column is marked with the third-order interatomic force constants (3rd-IFCs) for $\kappa_L$. 
\label{fig2}}
\end{figure*}

$\kappa_{3ph}^{HA}$, $\kappa_{3ph}^{SCPH}$, $\kappa_{3,4ph}^{SCPH}$ of Si$_4$O$_8$, Si$_4$O$_8$Kr, and Si$_4$O$_8$Cu versus temperature are showed in Fig.~\ref{fig1} (e).
This result indicates an anomalous increase in $\kappa_L$ after Kr doping and a decrease in $\kappa_L$ after Cu doping.
The influence of renormalization and 4ph scattering on $\kappa_L$ is elucidated 
through the ratios in the Fig.~\ref{fig2} (c). SCPH is the abbreviation of
self-consistent phonon with temperature-dependent phonon frequencies~\cite{wang2023role,xia2020high}. $\kappa_{3ph}^{SCPH}/\kappa_{3ph}^{HA}$ reflects the impact of phonon frequency ($\omega$) shifts due to temperature-dependent phonons. $\kappa_{3,4ph}^{SCPH}/\kappa_{3ph}^{HA}$ indicates the complete effect 
of 4th-order anharmonicity and $\kappa_{3,4ph}^{SCPH}/\kappa_{3ph}^{SCPH}$ reflects the additional influence
of 4ph scattering on top of the 3ph process.
In the Si$_4$O$_8$, Si$_4$O$_8$Kr and Si$_4$O$_8$Cu, the values of $\kappa_{3,4ph}^{SCPH}/\kappa_{3ph}^{SCPH}$ 
are 0.094, 0.141, and 0.147, respectively. This indicates that 4ph scattering significantly increases the phonon scattering. Moreover, all three materials exhibit strong phonon frequency shift effects, as evidenced by the 
relatively large values of $\kappa_{3ph}^{SCPH}/\kappa_{3ph}^{HA}$. 
Generally, 4ph scattering decreases $\kappa_L$ while the SCPH increases $\kappa_L$. In a practical situation, both
4ph scattering and phonon frequency shift compete in determining the final $\kappa_L$~\cite{wang2023role}.
The values of $\kappa_{3,4ph}^{SCPH}/\kappa_{3ph}^{HA}$ for Si$_4$O$_8$, Si$_4$O$_8$Kr, and Si$_4$O$_8$Cu are 
0.226, 0.153, and 0.289, indicating that 4ph scattering plays a more substantial role than the normalized 
phonon frequency shift in all three materials.  

The thermal conductivity spectrum $\kappa_L$($\omega$) and cumulative $\kappa_L$ at 300 K are shown in Fig.~\ref{fig1} (f). 
Low-frequency acoustic phonons contribute significantly to $\kappa_L$.
To elucidate the abnormal heat transport mechanism of doping, we obtain the phonon 
dispersions and phonon density of states (PDOS), as shown in Fig.~\ref{fig2} (a). Apparently, none of the materials 
exhibit imaginary frequencies in the harmonic approximation (HA), indicating the 
dynamic stability of these structures. The PDOS reveals that Si$_4$O$_8$ features a 
flat phonon mode in the mid-frequency range ($\omega$ = 120-150 cm$^{-1}$). However,
doping with the Kr atom eliminates the low-frequency flat mode. Cu atom doping introduces 
a rattling mode, particularly leading to significant flattening of the low-frequency 
phonon branch ($\omega$ = 20-50 cm$^{-1}$). These rattling modes increase the 
phonon-phonon scattering ~\cite{PhysRevLett.114.095501}.

The result indicates that the flat mode is primarily influenced by the vibrations of the filling atoms within the nanocages, 
and doping of the Cu atom plays an important role in the rattling modes of the atomic vibrations.
In clathrate structures, the extremely low $\kappa_L$ 
due to doping arises from the absence of avoided dispersion crossings of filling modes, significantly enhancing scattering channels~\cite{li2015ultralow, tse2005anharmonic, paschen2013thermopower, christensen2008avoided}.

Comparing the temperature-dependent phonon spectrum of Si$_4$O$_8$ at 0 K, 300 K, and 900 K, these three materials exhibit dynamical support stability as shown in Fig.~\ref{fig2} (b) and Fig. S4. There is an obvious phonon hardening phenomenon.

Firstly, we explore the reason for the decrease in $\kappa_L$ of Si$_4$O$_8$Cu. 
Fig. S5 (a) shows the calculated heat capacity (C$_V$) of Si$_4$O$_8$, Si$_4$O$_8$Kr, and Si$_4$O$_8$Cu as a function of temperature. 
It can be seen that the C$_V$ of Si$_4$O$_8$Cu is the highest, which is in contrast to the observed reduction in $\kappa_L$ of Si$_4$O$_8$Cu.
The change of phonon group velocities is small due to the doping, as shown in Fig. S5 (b).

\begin{figure*}
\includegraphics[width=0.7\textwidth]{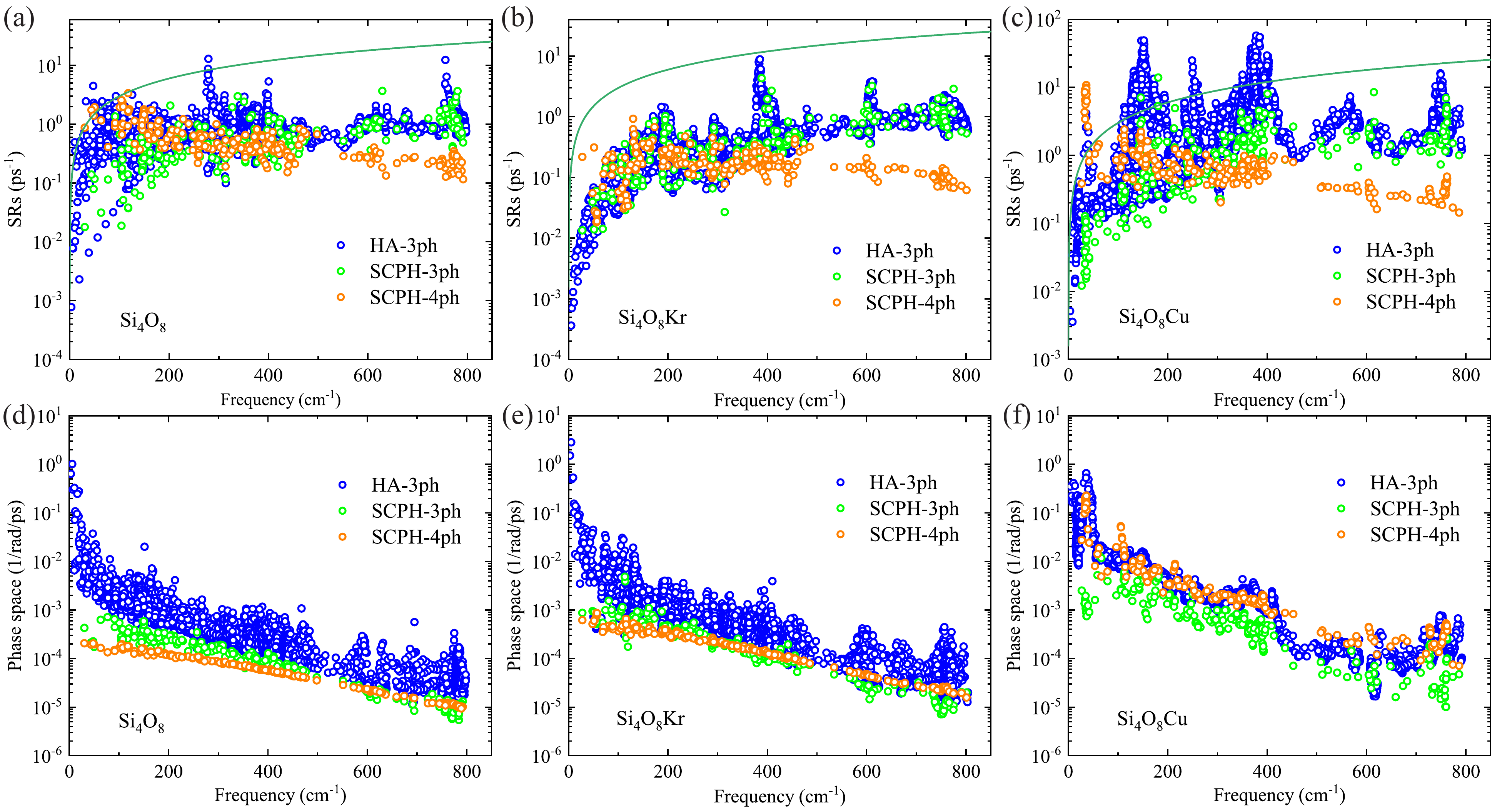}
\caption{
(a)--(c) Scattering ratios (SRs) and (d)--(f) weighted phase space as a function of frequency of Si$_4$O$_8$, Si$_4$O$_8$Kr and Si$_4$O$_8$Cu at 300 K. The blue, green, and orange colors represent 3ph scattering affected by the harmonic approximation (HA-3ph), 3ph scattering affected by SCPH (SCPH-3ph), and 4ph scattering affected by SCPH (SCPH-4ph). The solid green lines in (a)-(c) show that the scattering rate is equal to the phonon frequency ($1/\tau = \omega/2\pi$).
\label{fig3}}
\end{figure*}

Next, it is further considered that $\tau$ plays an important role in reducing $\kappa_L$ of Si$_4$O$_8$Cu. 
We have plotted the curve of $\tau$ as a function of phonon frequency, represented by $1/\tau = \omega/2\pi$. 
Once the single relaxation time exceeds the curve of $1/\tau = \omega/2\pi$, indicating that the $\tau$ is shorter than one vibrational period, 
the phonon quasiparticle picture is no longer valid~\cite{wang2023role}. As shown in Fig {\ref{fig3}} (a)-(c), the majority of 3ph and 4ph scattering events are distributed below the $1/\tau = \omega/2\pi$ curve, confirming the validity of the BTE solution in this study~\cite{PhysRevB.104.224304, xia2020particlelike,mukhopadhyay2018two}.

We interpret the difference in $\kappa_L$ among Si$_4$O$_8$, Si$_4$O$_8$Kr, and Si$_4$O$_8$Cu based on their phonon scattering rates (SRs), 
as shown in Fig.~\ref{fig3} (a)-(c) and Fig. S6. 
In Si$_4$O$_8$Cu, the scattering rates are nearly an order of magnitude higher than those in Si$_4$O$_8$ at most frequencies. Cu doping induces significantly higher scattering rates, leading to a reduction in $\kappa_L$~\cite{PhysRevLett.114.095501}. To further explore the potential mechanism behind the strong 4ph scattering, we analyzed the 3ph and 4ph scattering phase spaces of Si$_4$O$_8$, Si$_4$O$_8$Kr and Si$_4$O$_8$Cu at 300 K. The results indicate that the scattering phase space of Si$_4$O$_8$Kr increases upon doping with the Kr atom, while the Cu atom doping introduces rattling modes, causing a reduction in flat band frequencies. This significantly enhances the scattering channels and increases the scattering probability, as shown in Fig.~\ref{fig3} (e)-(f) and Fig. S7.

Interestingly, there is an anomalous increase in $\kappa_L$ upon doping with the Kr atom. Despite Si$_4$O$_8$ being an ultrahard material, a lattice strain of 0.927$\%$ was observed in Si$_4$O$_8$Kr, resulting in a $\kappa_L$ of 87.306 W m$^{-1}$ K$^{-1}$ at 300 K, as shown in Table 1. Subsequently, we calculated the $\kappa_L$ of Si$_4$O$_8$ when subjected to a 0.927 $\%$ strain, resulting in $\kappa_L$ of 93.788 W m$^{-1}$ K$^{-1}$ at 300 K, as displayed in Table 1. We observed that the $\kappa_L$ of strained Si$_4$O$_8$ is higher than that of Si$_4$O$_8$ doped with the Kr atom. Therefore, the enhanced $\kappa_L$ in Si$_4$O$_8$Kr primarily arises from the framework strain induced by the doping, rather than the rattling mode. This framework strain leads to an upward shift in the phonon spectrum, especially for the acoustic and low-frequency optical branches, as shown in Fig.~\ref{fig2} (a). Concurrently, Kr doping contributes to the reduction in $\kappa_L$.

\begin{table*}\scriptsize
        \parbox{1.0\textwidth}{
		\caption
		{The Influence of doped atomic mass on the change of lattice constant (a) and strain effect of Si$_4$O$_8$ and Si$_4$O$_8$X (X = Cu, Ar, Kr, and Xe).}
		 \label{tablabel1}}	
		\renewcommand\arraystretch{0.9} 
	\begin{tabular*}{1.0\textwidth}{p{0.5cm}*{6}{p{0.180\textwidth}}}
			\hline \hline
			& \centering Mass of X (u) & \centering a ({\AA}) & \centering Strain $\varepsilon$ (\%)  &  \centering $\kappa_L$ (W m$^{-1}$ K$^{-1}$) & \centering $\kappa_L$ / $\kappa_L$ (Si$_4$O$_8$) & \\ 
                \hline
		      Si$_4$O$_8$   & \centering 0 &  \centering 5.312 & \centering 0 &  \centering 28.289 & \centering 1 &\\  
                Si$_4$O$_8$   & \centering 0 & \centering 5.361 & \centering 0.927  &  \centering 93.788 & \centering 3.315 &\\    
                Si$_4$O$_8$Cu & \centering 63.55 & \centering 5.312 & \centering 0.007  &  \centering 17.648 & \centering 0.624 &\\
                Si$_4$O$_8$Ar & \centering 39.95 & \centering 5.343 & \centering 0.584  &   \centering 80.821 & \centering 2.857 &\\
                Si$_4$O$_8$Kr & \centering 83.80 & \centering 5.361 & \centering 0.927  &  \centering 87.312 & \centering3.112 &\\
                Si$_4$O$_8$Xe & \centering 131.29 & \centering 5.397 & \centering 1.600   &  \centering 121.405 & \centering 4.291 &\\
			\hline \hline
	\end{tabular*}
\end{table*}

Next, we compare the impact of harmonic and anharmonic contributions by exchanging the IFCs. Subsequently, the impact of lattice strain induced by exchanging harmonic and anharmonic IFCs on $\kappa_L$ is examined. 
As shown in Fig.~\ref{fig2} (d), the horizontal axis represents the input 2nd-IFCs, and annotations at the top of the bars indicate the input 3rd-IFCs, facilitating the computation of $\kappa_{3ph}^{HA}$.
Columns 1 and 4 share the same 3rd-IFCs, while their 2nd-IFCs differ. Exchanging force constants results in an increased $\kappa_L$, approaching $\kappa_L$ of Column 3 (Si$_4$O$_8$ $\varepsilon=0.927 \%$).
Therefore, under the influence of strain, harmonic effects emerge as the primary contributing factor to the increase in $\kappa_L$ upon doping of 2D silica.
Simultaneously, under the influence of strain induced by doping, both scattering rates and phase space decrease, corresponding to the observed increase in $\kappa_L$, as shown in Fig. S8-S9. The tensile strain effect on the $\kappa_L$ of 2D silica are shown in Fig. S10. 

By comparing the relative atomic mass of doping atoms in Table 1, it is evident that the 
doping atom X (X=Ar, Kr, and Xe) does not interact significantly with the host framework. 
The larger the mass of the doping atom, the greater the strain it induces. This phenomenon 
aligns with the behavior observed after applying the strain on the 2D silica~\cite{doi:10.1021/acsnano.6b05577}.
For instance, when the Cu atom is doped, it induces a framework strain of only 0.007 \%,
significantly smaller than the strain produced after doping with the Ar atom, which amounts 
to 0.584 \%. Despite copper (Cu) having a greater relative atomic mass than argon (Ar),
the strain of framework induced by Cu is even smaller than that induced by Ar. Therefore, 
it becomes necessary to consider the electronic interaction and distribution between the
doped Cu atom and the host framework.

\begin{figure*}
\includegraphics[width=0.7\textwidth]{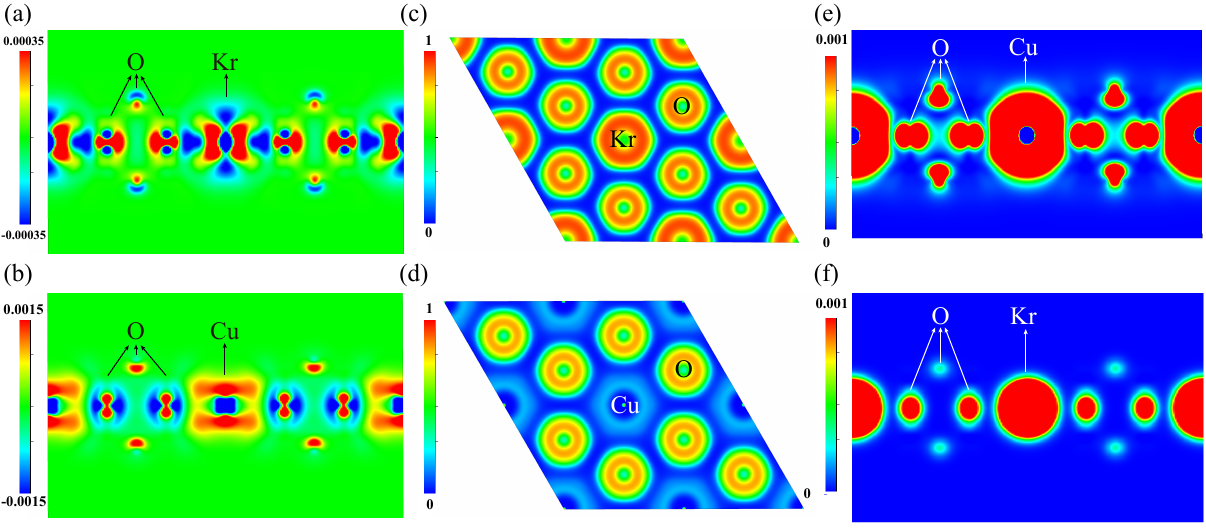}
\caption{
Effect of doping Kr and Cu atoms on the electron density of nanocages. (a)-(b) 2D projection of the differential electron density function on the (110) plane, the selection charge density ranges of Si$_4$O$_8$Kr and Si$_4$O$_8$Cu are $\pm$ 0.00035 and $\pm$ 0.0015 $e \cdot$ bohr$^{-3}$ respectively. 
(c)-(d) 2D projection of the electron local function (ELF) on the (001) plane
(e)-(f) 2D projection of the partial electron density function of Kr and Cu atoms in Si$_4$O$_8$. Selected partial charge densities ranging from 0 (blue) to 0.001 $e$ bohr$^{-3}$ (red).
\label{fig4}}
\end{figure*}

Thus, we conduct differential charge density calculations of Si$_4$O$_8$Kr and Si$_4$O$_8$Cu 
to investigate the change in electronic distribution. 
The regions with significant variations in the framework are attributed to the doping atom, On the one hand, since the Kr atom 
is in a chemical full-shell state and doping Kr will redistribute and squeeze the electron 
distribution of the framework, leading to the electrons moving to both sides in 
Fig.~\ref{fig4} (a). On the other hand, the Cu atoms have 
a strong attraction with the host framework, gathering the electrons toward the center in Fig.~\ref{fig4} (b).
Fig.~\ref{fig4} (c)-(d) display the electron localization function (ELF). In Si$_4$O$_8$Kr, the interaction between Kr atoms and the framework is relatively weak, leading to high electronic localization near the Kr atom. 
After the doping of Cu atoms, stronger interaction occurs between Cu atoms and the framework, resulting in smaller localization near Cu atoms. Combined with Bader charge analysis, it is shown that the doping of Cu and Kr atoms does not lead to the formation of chemical bonds with the host framework~\cite{PhysRevB.103.224101}.

Subsequently, partial electron density analysis is performed on the atomic character bands of Cu and Kr, as shown in Figure {\ref{fig4}} (e)-(f). By comparing the results of Si$_4$O$_8$Kr and Si$_4$O$_8$Cu, it is found that a noticeable enhancement in the interaction between Cu guest atoms and the host nanocages, indicating a certain level of electronic interaction. 
However, it should be noted that the doping atoms did not form chemical bonds with the host framework, thus inhibiting large strain on the framework. In the case of the noble gas atom Kr, the interaction with the host cage was significantly weaker, primarily due to the absence of electronic interaction between the guest atoms and the host nanocage electrons.

In summary, we have studied the effect of adsorbed noble gas and copper atoms on $\kappa_L$ of 2D silica nanocages based on temperature-dependent phonons through self-consistent phonon theory and fourth-order multiphonon scattering. The adsorption of noble gases results in an anomalous increase in $\kappa_L$, while the presence of Cu atoms leads to a reduction in $\kappa_L$. 
Adsorption of noble gas atoms on 2D silica results in tensile lattice strain within the nanocages, reducing phonon scattering rates and consequently leading to an increase in $\kappa_L$. In contrast to the weak interaction between noble gas atoms and the nanocages, Cu atom doping confines the distribution of electronic states without inducing significant strain in the nanocage, through its interaction with the nanocage electrons. The introduction of Cu atoms induces significant broadening and frequency shifts in the low-frequency phonon branches of the system, leading to the creation of large flat bands. This alteration, coupled with the enhancement of phonon scattering probability through rattling modes, results in a reduction of $\kappa_L$. 

Our results illustrate the microscopic heat transfer mechanism of adsorbed atoms in 2D silica nanocages, emphasizing the significant roles of lattice distortion and changes in electron density distribution in the $\kappa_L$. We also show the four-phonon scatterings play an indispensable role in the computation of $\kappa_L$ of 2D silica. Our study might inspire further theoretical and experimental investigations exploring materials with the kagome and glasslike $\kappa_L$~\cite{liu2023glass,tong2023glass,wang2023role}.

See the Supplementary Material for the phonon spectrum, the minimum energy path for doped Ar and Xe atoms, the renormalized phonon dispersion at different temperatures, heat capacity as a function of
temperature, phonon group velocity, Grünisen parameter, phonon scattering rates, phonon phase space and $\kappa_L$ of 2D silica with applied tensile strain.

The authors acknowledge the support from the National Natural Science Foundation of China (No.12104356 and No.52250191) and the Key Research and Development Program of the Ministry of Science and Technology (No.2023YFB4604100). 
Z.G. acknowledges the support of China Postdoctoral Science Foundation (No. 2022M712552), the Opening Project of Shanghai Key Laboratory of Special Artificial Microstructure Materials and Technology (No.Ammt2022B-1), and the Fundamental Research Funds for the Central 
Universities. We thank Prof. Mengyang Li for helping with the discussions.

\textbf{AUTHOR DECLARATIONS}

\textbf{Conflict of Interest}

The authors have no conflicts to disclose.

\textbf{DATA AVAILABILITY}

The data that support the findings of this study are available from the corresponding authors upon reasonable request.

\end{document}